\magnification=\magstep1 
\font\romsix=cmr6 scaled\magstep0  
\font\romeight=cmr8 scaled\magstep0  
\font\romnine=cmr9 scaled\magstep0  
\font\bigrom=cmr10 scaled\magstep2 
\def\ident#1{}
\def\goright{}
\def\Tr{\mathop{\rm Tr}\nolimits} 
\def\frac#1#2{{#1 \over #2}}
\def\half{{1\over2}}
\def\header#1{{
  \removelastskip\vskip 20pt plus 40pt \penalty-200 \vskip 0pt plus -32pt
  \NI\bf #1}\nobreak\medskip\nobreak}
\def\subheader#1{\removelastskip
 \vskip 10pt plus 20pt \penalty-200 \vskip 0pt plus -16pt
            \NI{\it #1}\nobreak\smallskip}
\def\standardpage{\hsize=6 true in \vsize= 8.5 true in \hoffset=0.25 true in}
\def\tenstart{\lineskiplimit=1pt\normalbase \tenrm ~~~~~~\par} 
\overfullrule=0pt
\def\newline{\hfil\break}
\def\negtwomu{\mskip -2mu}
\def\refmark{}
\def\eq#1{Eq.\ (#1)} 
\def\NI{\noindent}
\def\pritem#1{\vskip .04in\item{[#1]}}
\newcount\ftnum  \newcount\checknum  \newif\ifeqnerr
\def\aq#1{\global\advance\ftnum by 1 \xdef#1{\the\ftnum}}
\newcount\eftnum  \newcount\echecknum  \newif\ifeeqnerr
\def\eaq#1{\global\advance\eftnum by 1 \xdef#1{\the\eftnum}}
\def\newref#1{\aq#1#1}
\def\numeq#1{\eaq#1   \goright \ident#1\eqno(#1)$$} 
\def\nr{\newref}
\def\rk{\refmark}
\def\ph#1{\phi^{(#1)}}
\def\Sprim{S_{\hbox{\romsix prim}}}
\def\xs#1{x^{(#1)}}
\def\nus#1{\nu^{(#1)}}
\def\xsk{\xs{k}}\def\nusk{\nus{k}}
\def\delnu{\delta\negtwomu\nu}
\def\tauosc{\tau_{\hbox{\romsix osc}}}
\newread\epsffilein    
\newif\ifepsffileok    
\newif\ifepsfbbfound   
\newif\ifepsfverbose   
\newif\ifepsfdraft     
\newdimen\epsfxsize    
\newdimen\epsfysize    
\newdimen\epsftsize    
\newdimen\epsfrsize    
\newdimen\epsftmp      
\newdimen\pspoints     
\pspoints=1bp          
\epsfxsize=0pt         
\epsfysize=0pt         
\def\epsfbox#1{\global\def\epsfllx{72}\global\def\epsflly{72}%
   \global\def\epsfurx{540}\global\def\epsfury{720}%
   \def\lbracket{[}\def\testit{#1}\ifx\testit\lbracket
   \let\next=\epsfgetlitbb\else\let\next=\epsfnormal\fi\next{#1}}%
\def\epsfgetlitbb#1#2 #3 #4 #5]#6{\epsfgrab #2 #3 #4 #5 .\\%
   \epsfsetgraph{#6}}%
\def\epsfnormal#1{\epsfgetbb{#1}\epsfsetgraph{#1}}%
\def\epsfgetbb#1{%
\openin\epsffilein=#1
\ifeof\epsffilein\errmessage{I couldn't open #1, will ignore it}\else
   {\epsffileoktrue \chardef\other=12
    \def\do##1{\catcode`##1=\other}\dospecials \catcode`\ =10
    \loop
       \read\epsffilein to \epsffileline
       \ifeof\epsffilein\epsffileokfalse\else
          \expandafter\epsfaux\epsffileline:. \\%
       \fi
   \ifepsffileok\repeat
   \ifepsfbbfound\else
    \ifepsfverbose\message{No bounding box comment in #1; using defaults}\fi\fi
   }\closein\epsffilein\fi}%
\def\epsfclipoff{\def\epsfclipstring{\ifepsfdraft\space clip\fi}}%
\epsfclipoff
\def\epsfsetgraph#1{%
   \epsfrsize=\epsfury\pspoints
   \advance\epsfrsize by-\epsflly\pspoints
   \epsftsize=\epsfurx\pspoints
   \advance\epsftsize by-\epsfllx\pspoints
   \epsfxsize\epsfsize\epsftsize\epsfrsize
   \ifnum\epsfxsize=0 \ifnum\epsfysize=0
      \epsfxsize=\epsftsize \epsfysize=\epsfrsize
      \epsfrsize=0pt
     \else\epsftmp=\epsftsize \divide\epsftmp\epsfrsize
       \epsfxsize=\epsfysize \multiply\epsfxsize\epsftmp
       \multiply\epsftmp\epsfrsize \advance\epsftsize-\epsftmp
       \epsftmp=\epsfysize
       \loop \advance\epsftsize\epsftsize \divide\epsftmp 2
       \ifnum\epsftmp>0
          \ifnum\epsftsize<\epsfrsize\else
             \advance\epsftsize-\epsfrsize \advance\epsfxsize\epsftmp \fi
       \repeat
       \epsfrsize=0pt
     \fi
   \else \ifnum\epsfysize=0
     \epsftmp=\epsfrsize \divide\epsftmp\epsftsize
     \epsfysize=\epsfxsize \multiply\epsfysize\epsftmp   
     \multiply\epsftmp\epsftsize \advance\epsfrsize-\epsftmp
     \epsftmp=\epsfxsize
     \loop \advance\epsfrsize\epsfrsize \divide\epsftmp 2
     \ifnum\epsftmp>0
        \ifnum\epsfrsize<\epsftsize\else
           \advance\epsfrsize-\epsftsize \advance\epsfysize\epsftmp \fi
     \repeat
     \epsfrsize=0pt
    \else
     \epsfrsize=\epsfysize
    \fi
   \fi
   \ifepsfverbose\message{#1: width=\the\epsfxsize, height=\the\epsfysize}\fi
   \epsftmp=10\epsfxsize \divide\epsftmp\pspoints
   \vbox to\epsfysize{\vfil\hbox to\epsfxsize{%
      \ifnum\epsfrsize=0\relax
        \includegraphics{\ifepsfdraft}%
      \else
        \epsfrsize=10\epsfysize \divide\epsfrsize\pspoints
        \includegraphics{\ifepsfdraft}%
      \fi
      \hfil}}%
\global\epsfxsize=0pt\global\epsfysize=0pt}%
{\catcode`\%=12 \global\let\epsfpercent=
\long\def\epsfaux#1#2:#3\\{\ifx#1\epsfpercent
   \def\testit{#2}\ifx\testit\epsfbblit
      \epsfgrab #3 . . . \\%
      \epsffileokfalse
      \global\epsfbbfoundtrue
   \fi\else\ifx#1\par\else\epsffileokfalse\fi\fi}%
\def\epsfempty{}%
\def\epsfgrab #1 #2 #3 #4 #5\\{%
\global\def\epsfllx{#1}\ifx\epsfllx\epsfempty
      \epsfgrab #2 #3 #4 #5 .\\\else
   \global\def\epsflly{#2}%
   \global\def\epsfurx{#3}\global\def\epsfury{#4}\fi}%
\def\epsfsize#1#2{\epsfxsize}
\let\epsffile=\epsfbox
\def\lsfig#1#2#3{\epsfxsize=#1in \epsfysize=#2in \epsffile{#3.eps}}
\def\normalbase{\baselineskip 12pt} \standardpage

\footline={\ifnum\pageno=1 \hfill \else\hss\tenrm\folio\hss\fi}
\tenstart

{\par\baselineskip 9 pt \romeight \hangindent 3 true in \hangafter -5 \NI
For the proceedings of the conference, ``Time's Arrows, Quantum Measurements and Superluminal Behavior," Naples, Italy, October 2000. To be published by the Italian CNR. \par}

\vskip .4 true in

\centerline{\bigrom Causality is an effect}


\bigskip
\centerline {L. S. Schulman}
\medskip
\centerline {\romnine Physics Department, Clarkson University}
\centerline {\romnine Potsdam, NY 13699-5820 USA}
\centerline{\romnine email: schulman@clarkson.edu}
\bigskip

{\narrower \romnine \baselineskip 10pt \def\it{\italnine} \def\bf{\boldnine}
\centerline{ABSTRACT} \vskip 2pt
Using symmetric boundary conditions at separated times, I show analytically that both the time ordering of (macroscopic) causality and the direction of entropy increase follow from these boundary conditions. In particular, when the endpoints have low entropy, these arrows of time point away from the ends and toward the middle. Causality in this context means that when perturbations are applied, the effect of the perturbation---the macroscopic change in the system's behavior---is confined to one temporal side of the perturbations. These results hold for both mixing and integrable systems, although relaxation for integrable systems is incomplete. Simulations are presented for purposes of illustration.
\par}

\header{1. Introduction}

By ``causality" I mean that if a system is perturbed the macroscopic effect occurs subsequent to the perturbation. There is a lot of baggage in this definition. First, I am not talking about the microscopic causality of relativistic quantum field theory, which is a statement about the vanishing of commutators (or anticommutators) at spacelike separations. Second, I am trying to avoid the many and subtle definitions that have appeared in the philosophical literature, some of which are close to mine, some of which are not. Then there is the word ``perturbation," which suggests a kind of control or free will. Finally, there is the term ``macroscopic," equivalent to a notion of coarse grains, yet another nontrivial concept.

Defining causality in terms of sequential order emphasizes its relation to the thermodynamic arrow of time. Indeed, some consider causality (with similar meaning and baggage) to be the primary concept [\nr\newton\rk, pp.\ 163--164], with other kinds of ordering (in particular, the second law of thermodynamics) consequences of it.

I will take neither of these concepts to be primary, and will instead derive both from a model, or caricature, of the expansion of the universe. This follows the ideas of Gold [\nr\gold\rk] and my own elaboration of them [\nr\correlating,\nr\timebook\rk], in particular emphasizing the notion of two-time boundary conditions. It is clear that it would be pointless to study causality as defined above using {\it initial\/} values for macroscopic problems, since such a formulation {\it forces} the effect of a perturbation to be subsequent. So in studying causality, as in studying the arrow of time, one should formulate the problem time-symmetrically if one's conclusions are to be noncircular.

I will find that both macroscopic causality and the second law, meaning entropy increase, can be {\it derived\/} in the appropriate two-time boundary condition context. For sufficiently chaotic dynamical systems both features flow naturally from the formalism. For integrable systems, relaxation can be imposed by averaging over frequencies. But with future conditioning additional time scales enter, and while one can still get relaxation and causality, there is not the same simplicity as for chaotic systems.

In Sec.\ 2 I introduce the general context for this discussion as well as notation. In the following section there is an analytic derivation of symmetric entropy increase for systems having appropriate two-time boundary conditions. Causality, treated in Sec.~4, is established using the same methods. In the last section numerical work is shown to illustrate the results of the previous sections. There are two appendices. In the first I give a general derivation of entropy increase when coarse graining is implemented at each time step. This is a master equation approach and is mainly pedagogical. In the second appendix I indicate how the notion of ``perturbation" need not depend on philosophical questions concerning free will.

\header{2. Framework and notation}

As in previous publications \aq \opposite \aq \oppvail
[\correlating--\oppvail\rk], my context is a two time boundary value problem in which macroscopic data are specified at an early time, ``0," and a late time, ``$T$." At {\it both\/} boundary times the system is in a restricted state (i.e., low entropy). For the systems previously studied the entropy increases, with varying degrees of monotonicity, away from both boundary times. Moreover, for chaotic systems, if the interval between the boundary times exceeds twice the system's relaxation time, the initial relaxation is macroscopically indistinguishable from {\it un}conditioned time evolution. Furthermore, the evolution away from the final point (i.e., from $T$ to smaller values of the parameter $t$) is the symmetric image of the initial relaxation---this assertion is true even with conditions on the time evolution that are weaker than time reversal invariance. All these features have been used to argue for Gold's thesis. In some of the references above I have elaborated on my rationale for taking this approach, and will not repeat the argument here. Most of my previous demonstrations have used the cat map [\nr\arnold\rk] as the dynamical system, and computer simulations to provide the evidence.

In this article I will argue more generally, extending both the systems studied and the method of justification. In effect this explains why the simulations work, although a discussion without explicit equations occurs in [\timebook\rk] and embodies the essential ideas to be presented below.

Consider classical mechanics on a (phase) space $\Omega$. Let $\mu$ be the measure on $\Omega$ and $\mu(\Omega)=1$. Let the dynamics be given by a measure-preserving map $\ph{t}$ on $\Omega$, with $\ph{t}(\omega)$ the time-$t$ image of an initial point $\omega\in\Omega$. The time parameter, $t$, may be either continuous or discrete.

The notion of ``macroscopic" is provided by a coarse graining on $\Omega$. This is a finite set of sets of strictly positive measure that cover $\Omega$: $\{\Delta_\alpha\}$, $\alpha=1,\ldots,G$, with $\cup_\alpha\Delta_\alpha=\Omega$, $\Delta_\alpha\cap\Delta_\beta = \emptyset$ for $\alpha\neq\beta$. Let $\chi_\alpha$ be the characteristic function of $\Delta_\alpha$ and let $v_\alpha=\mu(\Delta_\alpha)$. If $f$ is a function on $\Omega$, its coarse graining is defined to be
$$\widehat f(\omega) 
     \equiv \sum_\alpha \frac{\chi_\alpha(\omega)}{v_\alpha} f_\alpha
      = \sum_\alpha \chi_\alpha \langle f\rangle_\alpha    \,,
        \quad \hbox{with~}
   f_\alpha =\!\int\! d\mu\,\chi_\alpha(\omega) f(\omega)
         \;,\
   \langle f\rangle_\alpha \equiv \frac{f_\alpha}{v_\alpha}
    \,.\numeq{\coarse}
Thus $\langle f\rangle_\alpha$ is the average of $f$ on $\Delta_\alpha$ and $\int_\Omega f = \int_\Omega\widehat f$.

Let the system's distribution in $\Omega$ be described by a density function $\rho(\omega)$. One can think of this distribution in more than one way. In terms of the ideal gas of cat map atoms that I have used before, $\rho$ can be thought of as the density of atoms on the phase space, $I^2$, on which a single cat map lives. (In this case $\rho$ is a sum of $\delta$-functions.) More generally, one can think of $\Omega$ as the $6N$-dimensional phase space of $N$ particles in three space dimensions. In this way there is no restriction in allowing interactions among the particles.

If one takes a primitive notion of entropy as 
$$\Sprim=-\int_\Omega \rho(\omega) \log(\rho(\omega))\,d\mu\,,$$
then $\Sprim$ is constant in time, trivially by virtue of the measure preserving property of $\ph{t}$ (and its invertibility). The entropy that I will use for studying irreversibility involves coarse graining and is defined as
$$ S(\rho)=\Sprim(\widehat\rho)
         =-\int_\Omega \widehat\rho \log\widehat\rho\,d\mu\,.\numeq{\entdef}
It is easy to show that
$$S(\rho)=S(\rho_\alpha|v_\alpha)\,,$$
where, as in \eq{\coarse}, $\rho_\alpha = \int_{\Delta_\alpha}\rho\,d\mu$, and the function $S(p|q)$ is the {\it relative entropy} defined by
$$S(p|q) \equiv -\sum_x p(x) \log\left(\frac{p(x)}{q(x)}\right)\,,$$ 
with $p$ and $q$ probability distributions such that $p(x)$ vanishes only if $q(x)$ does. Note that $\sum \rho_\alpha =\int \rho=1$, and that all $v_\alpha$ are nonzero [\nr\differentdef\rk].

\header{3. Time-dependence of the entropy, with and without future conditioning}

The system is required to start ($t=0$) in a subset $\epsilon_0\subset\Omega$ and end ($t=T$) in a subset $\epsilon_T\subset\Omega$. The points of $\Omega$ satisfying this two-time boundary condition are
$$ \epsilon=\epsilon_0 \cap \ph{-T}(\epsilon_T)\,.\numeq{\twotimeset}
The set $\epsilon$ can be empty. I have argued though [\nr\accuracy,\timebook\rk] that for chaotic dynamics and for sufficiently long times $T$ there exist solutions, i.e., $\epsilon\neq\emptyset$. Moreover, for such times 
$$\mu(\epsilon)\sim \mu(\epsilon_0) \mu(\epsilon_T)\,.  \numeq{\productmeasure}
To see how this comes about, consider mixing dynamics. The map $\ph{t}$ is mixing if
$$ \lim_{t\to\infty} \mu\left(A\cap\ph{t}(B)\right)=\mu(A)\mu(B)   \numeq{\mixing}
for measurable subsets $A$ and $B$ of $\Omega$. For such systems \eq{\productmeasure} will be satisfied in the $t\to\infty$ limit. This limit says nothing about rates of convergence, but I will assume that there is some time $\tau$ such that the decorrelation condition (\eq{\mixing}) holds to good accuracy for $t\geq \tau$ [\nr\goodaccuracy\rk]. The set $\epsilon$ will therefore be nonempty for $t\geq\tau$. Under $\ph{t}$, $\epsilon$ becomes
$$\epsilon(t)=\ph{t}(\epsilon_0)\cap\ph{t-T}(\epsilon_T)\,.$$
To calculate the entropy, the density, which was $\rho(0)= \chi_\epsilon/\mu(\epsilon)$ at time-0, must be coarse grained. The important quantity for the entropy calculation is 
$$\rho_\alpha(t)=\frac{\mu\left(\Delta_\alpha\cap\epsilon(t)\right)}
     {\mu(\epsilon)}
=\frac{\mu\left(\Delta_\alpha\cap\ph{t}(\epsilon_0)\cap\ph{t-T}(\epsilon_T)\right)}
      {\mu(\epsilon)} \,.$$
If $T-t>\tau$ then the following will hold
$$\eqalign{
\mu\left(\Delta_\alpha\cap\ph{t}(\epsilon_0)\cap\ph{t-T}(\epsilon_T)\right)
=&\mu\left(\Delta_\alpha\cap\ph{t}(\epsilon_0)\right)
       \mu\left(\ph{t-T}(\epsilon_T)\right)                           \,,   \cr
\mu(\epsilon)=&\mu(\epsilon_0)\mu\left(\ph{-T}(\epsilon_T)\right)     \,.   \cr
}$$
Using the measure-preserving property of $\ph{t}$, the factors $\mu(\epsilon_T)$ in both numerator and denominator of $\rho_\alpha$ cancel, leading to
$$\rho_\alpha=\mu\left(\Delta_\alpha\cap\ph{t}(\epsilon_0)\right)\Big/\mu(\epsilon_0) \,.$$
This is precisely what one gets {\it without\/} future conditioning, so that all macroscopic quantities, and in particular the entropy, are indistinguishable from their unconditioned values.

Working backward from time-$T$ one obtains an analogous result. Define a variable $s\equiv T-t$ and set $\tilde\epsilon(s) \equiv\epsilon(T-s)$. Then
$$\tilde\epsilon(s)=\ph{T-s}(\epsilon_0)\cap\ph{-s}(\epsilon_T) \,.$$
If $s$ satisfies $T-s>\tau$, then when the density associated with $\tilde\epsilon(s)$ is calculated, its dependence on $\epsilon_0$ will drop out. It follows that
$$\rho_\alpha(s)=\mu\left(\ph{-s}(\epsilon_T)\right)\Big/\mu(\epsilon_T) \,.$$
For a time-reversal invariant dynamics this will give the entropy the same time dependence coming back from $T$ as going forward from 0. It is interesting that the cat map is not strictly time-reversal invariant (by definitions of the form given in [\nr\unstable\rk]) but, as I have shown repeatedly, its entropy as a function of time is symmetric. The reason is that the Lyapunov exponent is the same for the map and its inverse. For the cat map, there isn't much choice: the $2\times2$ matrix has only two eigenvalues and their product is unity. But I expect the similarity of macroscopic dynamics in both directions to obtain even for richer systems. Thus, comparing true physical dynamics with its time-reversed counterpart, ordinary macroscopic relaxation should be the same, yielding symmetric entropy dependence. I justify this expectation by the absence (so far) of any time-reversal or CP violating observations at the atomic level, as well as the assumption that ordinary physical relaxation processes, accounting for the thermodynamic arrow of our experience, occur at grosser levels than those at which CP violation has been detected.

It is worth putting into words the essence of the mathematical argument just given. The set $\epsilon$ is a subset of $\epsilon_0$; which points of $\epsilon_0$ are also in $\epsilon$ is determined by the choppy characteristic function of the set $\ph{-T}(\epsilon_T)$. For long enough times, $T$, the good points of $\epsilon$ are Poisson distributed within $\epsilon_0$ [\timebook\rk]. Thus following $\epsilon$ forward in time (with $\phi$) is like following a random subset of $\epsilon_0$. But such time evolution is one way of studying $\epsilon_0$ itself. If you wanted to do a Monte Carlo study for the evolution of $\epsilon_0$, your technique would be to follow the time dependence of a random subset. The pseudo-randomness imposed by the characteristic function of $\ph{-T}(\epsilon_T)$ is not worse than other kinds of pseudo-randomness.
 
The same pseudo-randomness holds for $\epsilon(t)$ ($t>0$), provided the time to the final point, $T-t$, is greater than $\tau$. I have used the mixing property to argue for randomness, but I expect weaker conditions of ergodicity to be sufficient in physically relevant situations.

\subheader{Integrable systems (relaxation)}

Without mixing or some kind of ergodicity the foregoing arguments fail. However, harmonic oscillators can be quite useful in studies of relaxation [\nr\huerta\rk], although in previous two-time boundary value studies [\correlating\rk] deficiencies were noted. The general idea is that although an individual oscillator does not spread in phase space, if enough different frequencies are taken there is relaxation.

Rather than work with sets, as above, I consider an ``ideal gas" of $N$ oscillators, with oscillator \#$k$ having position $\xsk$ and frequency $\nusk$. (For convenience I take the period of these oscillators to be 1 (rather than $2\pi$) and use frequency rather than angular frequency.) Time evolution is given by $\xsk(t)=\xsk_0+\nusk t$ (mod~1). The boundary conditions are
$$\eqalign{
&0\leq \xsk_0\leq\delta x \hbox{~~\&~} 0\leq \xsk(T)\leq\delta x \,,
          \hbox{~~with~} \xsk(T)=\xsk_0+\nusk T  ~(\hbox{mod 1}) \,,            \cr
&\nu_0\leq\nusk\leq\nu_0+\delnu  \qquad\hbox{($\nu$ does not change in time)}  \,.\cr
}\numeq{\osccond}
Both $\xsk$ and $\nusk$ can be randomly selected consistent with these conditions. From the final-time condition on $x$ it follows that for sufficiently large $T$ there is a nonempty finite set of integers $\{n_\ell\}$ so that
$$\frac{-\xsk_0}T \leq \nusk - \frac{n_\ell}T \leq \frac{\delta x-\xsk_0}T 
    \,,   \numeq{\nucondition}
with $\nusk$ within the permitted range. One can plot the set of allowed initial points in the $x$-$\nu$ plane. Within the rectangle $[0,\delta x] \times [\nu_0,\nu_0+\delnu]$
the solution points fall in a sequence of parallel parallelograms. Each is bounded by vertical lines at $x=0,\delta x$ and by lines of slope $-1/T$ defining the upper and lower values for each $n_\ell$ (except that parallelograms going outside $[\nu_0,\nu_0+\delnu]$ are cut off). For large $T$ there will be many such parallelograms; for small $T$, few or none.

A natural coarse graining is to divide the $x$-range, $[0,1]$, into $G$ intervals and look only at values of $x$ to compute entropy, since $\nu$ does not change. For present purposes the boundary value quantity, $\delta x$, will be taken smaller than $1/G$.

I first examine the equilibration of this system without future conditions. Initially all points are in $[0,\delta x]$; they separate from one another only by virtue of possessing different frequencies. Equilibration is marked by $\delnu\, t \approx 1$, leading to the definition of a relaxation time $\tauosc= 1/\delnu$. It follows that on a time scale of $\tauosc$ the entropy will rise from $-\log G$ to 0 (with the stated condition, $\delta x<1/G$).

Now consider the situation with the future conditioning of \eq{\osccond}. For each $n_\ell$ of \eq{\nucondition}, the points of its parallelogram either remain in a single grain ($\forall t\leq T$) or at worst overlap two. This does not contribute significantly to entropy increase. Rather, the separation of individual parallelograms is required for equilibration. The number of such parallelograms is estimated by replacing $\nu$ by $n/T$ in the second part of \eq{\osccond}. It follows that the number of $n$ values is $T\delnu$. Neglecting grain overlap, this implies that the maximum entropy for the two-time boundary value problem is $\log (T\delnu) -\log G$. Therefore the system cannot fully relax unless $T>G/\delnu$. This is much longer than the unconditioned relaxation time, $\tauosc = 1/\delnu$, and is in sharp contrast to the behavior of mixing systems---for the cat map the ``$T$" necessary for normal initial relaxation is only twice the usual relaxation time.

There are further defects in the relaxation. Consider what happens when $t=T/2$. Write $\nu=(n+\beta)/T$, with $\beta$ a number on the order of $\delta x$ (cf.\ \eq{\nucondition}). Then $\xsk(T/2) = \xsk_0+n/2+\beta/2$. For even $n$ this means that many of the points are back in the original interval, or close to it. Thus the entropy will drop. The same happens, but less dramatically, for other divisors. Finally, there is an inherent weakness in {\it any} oscillator equilibrium, in that only half the dynamical variables relax at all---the frequencies ($\nu$) do not change.

In terms of the big bang-big crunch cosmological model considered in 
[\timebook--\oppvail\rk] these defects are probably irrelevant. It appears that the ``oscillators" in our cosmos are mostly the degrees of freedom of the electromagnetic field. These reach equilibrium through being coupled to massive matter, which presumably {\it does} relax appropriately. If they do satisfy a two-time boundary condition with, say, the boundary times at the decoupling epoch and its pre-big crunch partner, then I would not expect the timing to be so precise and coincident that one would get photon entropy-lowering at the cosmological midpoint (as in our ``even $n$" condition above). Moreover, photons do {\it not\/} equilibrate very well: witness the preservation of indications of spatial structure as deduced from cosmic background radiation. On the other hand, the spectrum of this radiation corresponds very well to equilibrium, the reason being interaction with matter prior to decoupling.

\header{4. Causality and peturbations}

The notion of macroscopic causality used here involves a perturbation. One imposes two-time boundary conditions and considers dynamical evolution with both unperturbed and perturbed dynamics. When solving the same boundary value problem, these rules will select different microscopic solutions. Although I will consider perturbations occurring only at a single moment in time, the microscopic solutions will (in general) differ everywhere. But it is the macroscopic solutions that allow a notion of causality. In principle macroscopic behavior could also differ at all times (except for the boundaries), but in a system with causality they will differ on only one side of the perturbation. For the usual causality, they will differ only after the perturbation. But we will also find that they can differ only before, where in this sentence and the last the words ``before" and ``after" are defined with respect to a microscopic time parameter, that, as will be seen, may differ from the natural thermodynamic time.

There is a delicate point here that is discussed in Appendix \hbox{B}. The term ``perturbation" suggests free will, while two-time boundary conditions sound like the opposite. Resolving issues of free will is not my objective, and the appendix is devoted to formulating the concept of perturbation in a purely physical context.

Although I will later give examples (figures) in terms of discrete time, for formal purposes it is easiest to work in continuous time and to imagine that the perturbation is instantaneous. The time interval for the boundary value problem is $[0,T]$. Call the unperturbed system A; its history, time evolution, dynamics and boundary conditions are exactly as described in the previous section. That is, it evolves under $\ph{t}$, its boundary conditions are $\epsilon_0$ and $\epsilon_T$, and its microstates are in the set
$$ \epsilon_A=\epsilon_0 \cap \ph{-T}(\epsilon_T)\numeq{\Aset}
(formerly called $\epsilon$). System B, the perturbed case, has an additional transformation act on it at time-$t_0$. Call this transformation $\psi$. It should not be dissipative---I do not want the arrow to arise from such an asymmetry alone [\nr\mixingpaper\rk]. $\psi$ is thus invertible and measure preserving. Successful solutions must go from $\epsilon_0$ to $\epsilon_T$ under the transformation $\ph{T-t_0}\psi\ph{t_0}$. The microstates for system B are therefore in
$$ \epsilon_B=\epsilon_0 \cap \ph{-t_0}\psi^{-1}\ph{-T+t_0}(\epsilon_T)\numeq{\Bset}

Clearly, $\epsilon_A$ and $\epsilon_B$ are different. But as I shall now show, for mixing dynamics and for sufficiently large $T$, the following hold: 1) for $t_0$ close to 0, the only differences in macroscopic behavior between A and B are for $t>t_0$;~~2) for $t_0$ close to $T$, the only differences in macroscopic behavior between A and B are for $t<t_0$. This means (recalling Sec.\ 3) that the direction of causality follows the direction of entropy increase.

The proof is nearly the same as that of the previous section. Again we use the time $\tau$ such that the mixing decorrelation holds for time intervals longer than $\tau$. First consider $t_0$ close to 0. The observable macroscopic quantities are the densities in grain-$\Delta_\alpha$, which are, for $t<t_0$,
$$\eqalign{
\rho_\alpha^{A}(t)
    &=\mu\left(\Delta_\alpha\cap\ph{t}(\epsilon_0)\cap\ph{t-T}(\epsilon_T)\right)
      \Big/\mu(\epsilon_A) \,,
\cr
\rho_\alpha^{B}(t)
    &=\mu\left(\Delta_\alpha\cap\ph{t}(\epsilon_0)\cap
           \left[\ph{t-t_0}\psi^{-1}\ph{t_0-T}\right](\epsilon_T)\right)
      \Big/\mu(\epsilon_B) \,.
\cr}$$
As before, the mixing property, for $T-t>\tau$, yields $\rho_\alpha^{A}(t)= \mu\left( \Delta_\alpha \cap \ph{t}(\epsilon_0) \right)/\mu(\epsilon_0)$, which is the initial-value-only macroscopic time evolution. For $\rho_\alpha^{B}$, the only difference is to add a step, $\psi^{-1}$. Unless $\psi^{-1}$ is diabolically contrived to undo $\ph{-u}$ for large $u$, this will not affect the argument that showed that the dependence on $\epsilon_T$ disappears. Thus A and B have the same macrostates before $t_0$.

For $t>t_0$, $\rho_\alpha^{A}(t)$ continues its behavior as before. For $\rho_\alpha^{B}(t)$ things are different:
$$\rho_\alpha^{B}(t)
  =\mu\left(\Delta_\alpha \cap \left[\ph{t-t_0}\psi\ph{t_0}\right](\epsilon_0)\cap
           \ph{t-T}(\epsilon_T)\right)
      \Big/\mu(\epsilon_B) \,~~~~(t>t_0).
$$
Now I require $T-t>\tau$. If this is satisfied the $\epsilon_T$ dependence drops out and
$$\rho_\alpha^{B}(t)
=\mu\left(\Delta_\alpha \cap \left[\ph{t-t_0}\psi\ph{t_0}\right](\epsilon_0)\right)
    \Big/\mu(\epsilon_0) \,.
$$
The shows that the effect of $\psi$ is the usual initial-conditions-only phenomenon.

If we repeat these arguments for $t$ such that $T-t$ is small, then just as we showed in Sec.\ 3, the effect of $\psi$ will only be at times $t$ {\it less than} $t_0$.

This manifestation of causality has a clear intuitive origin. As the perturbation time, $t_0$, choose a value small enough that the system has {\it not} equilibrated [\nr\onlymeaningful]. All points, $a\in\epsilon_A$ and $b\in\epsilon_B$, start in $\epsilon_0$ and end in $\epsilon_T$. Working backward from $t_0$, what can $b$ do? It must get to $\epsilon_0$. Since $t_0$ is less than the relaxation time, the places it can be are essentially the same places that $a$ can be. However, after the perturbation the need to arrive in $\epsilon_T$ places no macroscopic restriction on $b$, because from any coarse grain in $\Omega$ you can find your way into $\epsilon_T$. This is precisely because working {\it backward\/} from $T$, the set $\epsilon_T$ spreads throughout $\Omega$ in time $T-t_0$ (and in particular $\ph{t_0-T}\epsilon_T$ enters the coarse grain into which $\psi$ would send $b$ if there were no future conditioning). Thus, satisfying the changed boundary conditions is accomplished by keeping $a$ and $b$ close to one another {\it before\/} $t_0$, and allowing the perturbation a free hand in moving $b$ away from $a$, after $t_0$.

\subheader{Integrable systems (causality)}

As before, without mixing or some kind of ergodicity our arguments fail. Nevertheless, just as frequency smearing gave relaxation, however imperfect, it can give causality. Again an extended time scale is needed, but the intuitive reasoning just given continues to hold in the integrable case as well.

Consider a particular example, an oscillator of the sort discussed in Sec.\ 3. Take $\delta x$ so small that the condition in \eq{\osccond} forces all the points to have essentially the time dependence, $x=\nu t$, with $\nu T=n$. The angular frequency $\nu$ therefore satisfies $\nu=n/T$, with $n$ selected so that $\nu_0 \leq\nu \leq\nu_0 + \delnu$; for large $T$, this allows an extensive range of $n$ values. Now consider the following perturbation: at the moment $t_0$, $x$ is displaced by a macroscopic angle $\gamma$, i.e., $\gamma>1/G$. Solving the same boundary value problem gives $x=\nu t$ before $t_0$, and $x=\nu t+\gamma$ after $t_0$. With the perturbation, $\nu$ must satisfy $\nu=n/T -\gamma/T$, again yielding an extensive range of $n$ values for large $T$. The difference between the two ranges of $n$ values is $\gamma/T$, which for large enough $T$ will be a small fraction of all $n$ values that are common to the perturbed and unperturbed motion. Such $n$ are henceforth dropped from consideration.

For $n$ values that are common to the two solution sets, the difference between solutions with the same $n$ arises from the $\gamma$-dependent difference in $\nu$:
$$\hbox{Effect}_n=x_{\hbox{\romsix unperturbed}}-x_{\hbox{\romsix perturbed}} 
    = \cases{                \gamma t/T  & $t<t_0$                \cr
                            -\left(1-t/T\right)\gamma~~~ & $t>t_0$\cr}\,, \numeq{\effect}
which is independent of $n$. It follows that there is a difference between perturbed and unperturbed motion that is of order $\gamma$ through most of the time period $[0,T]$. The effect of the perturbation is felt both before and after. It thus appears that there is no causality, but closer consideration shows this conclusion to be wrong.

Recall that it is only meaningful to consider perturbations that take place before the system has relaxed (or close enough to $T$ that the reverse process has commenced). Thus the perturbation should occur for $t_0 <\tauosc = 1/\delnu$. On the other hand, for full relaxation the value of $T$ should be greater than $G/\delnu$, as discussed in Sec.~3. From \eq{\effect} the maximum value of the precursor---the noncausal term---is $t_0/T$, just before the perturbation. These considerations are combined to yield
$$ \frac1\gamma\cdot\hbox{Maximum noncausal precursor}\
        = \frac{t_0}T <\frac{\tauosc}{G/\delnu}
        =\frac1G       \,.  $$
But the size of a coarse grain is $1/G$, so that this precursor is in fact microscopic.

\subheader{Double arrow systems}

In [\opposite\rk] I showed that causality obtains in opposite directions in systems containing opposite arrows. The general principle is the same as that presented here although a detailed presentation would be more complicated by virtue of the simultaneous presence of two directions for causality. I will not provide an analytic demonstration and only mention this matter here for completeness.

\header{5. Numerical illustrations}

Although the purpose of the present article is to go beyond the numerical simulations of previous publications, I will illustrate the phenomena studied here.

\topinsert
\begingroup 
\def\moveleftstart{-.2truein} 
\def\downskip{-.4truein} 
\def\moveleftmid{-.01truein} 
\def\figsize{1.8}
~\vskip \downskip
\hbox{\hskip\moveleftstart
\lsfig{\figsize}{\figsize}{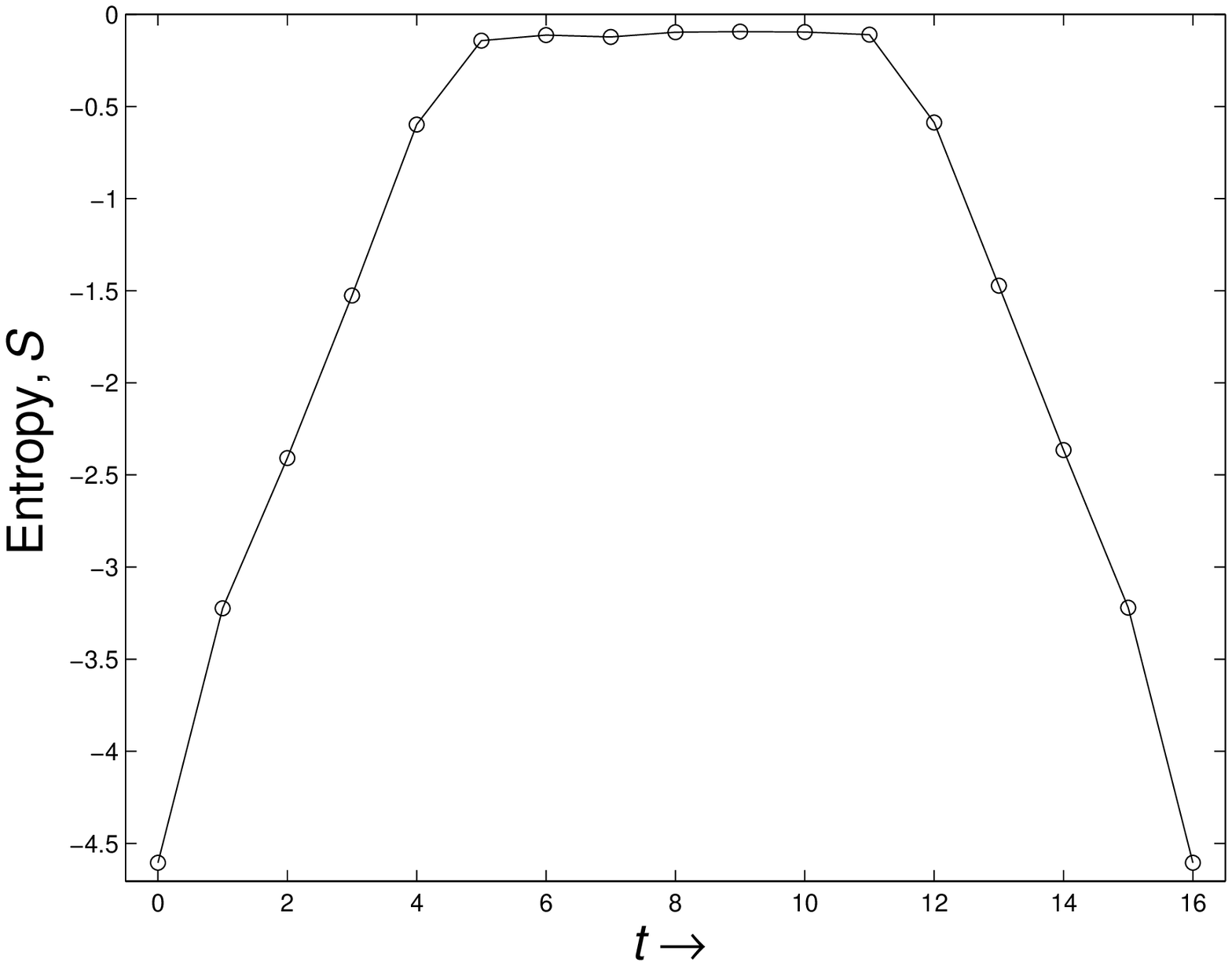}      \hskip \moveleftmid
\lsfig{\figsize}{\figsize}{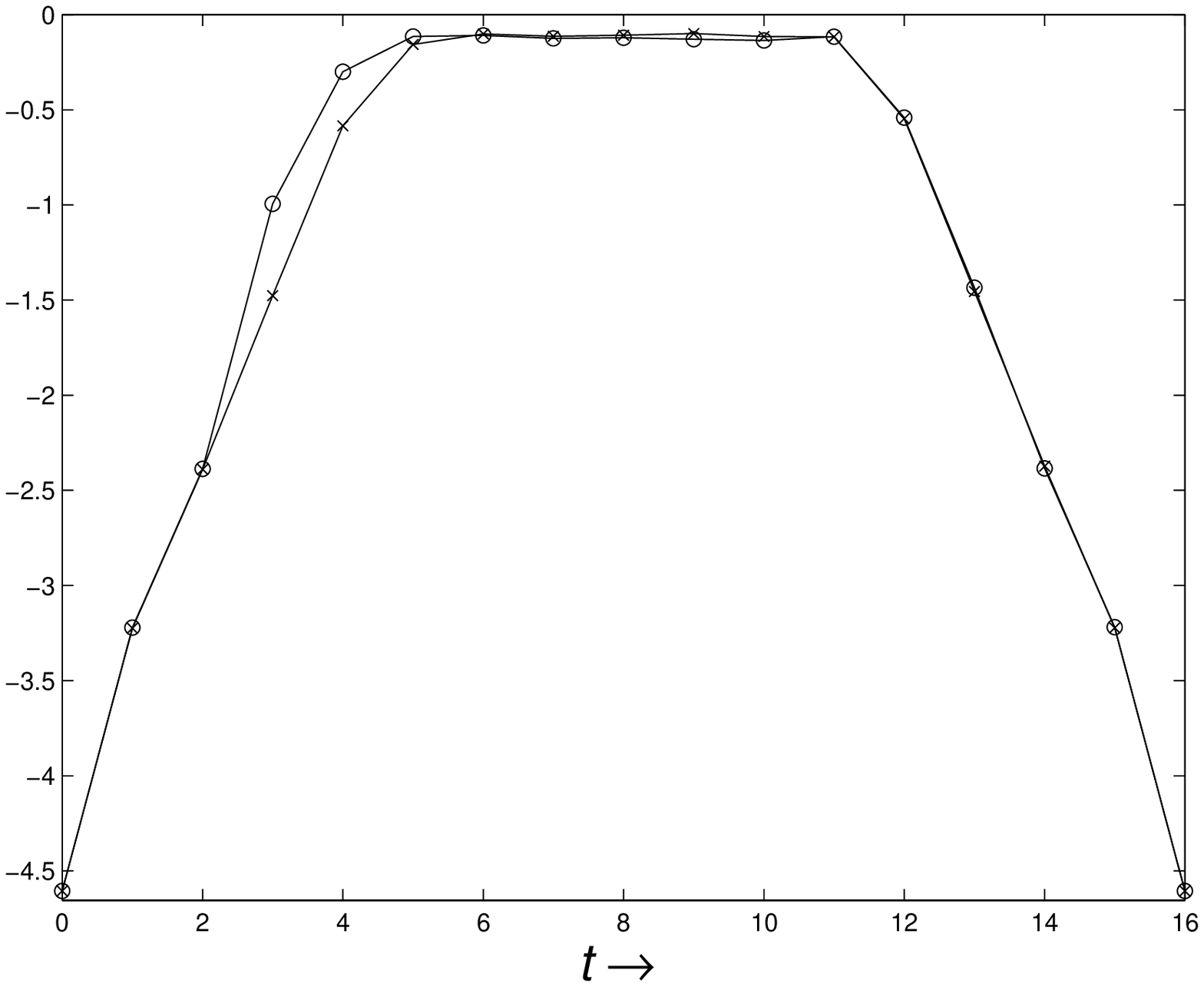}   \hskip \moveleftmid
\lsfig{\figsize}{\figsize}{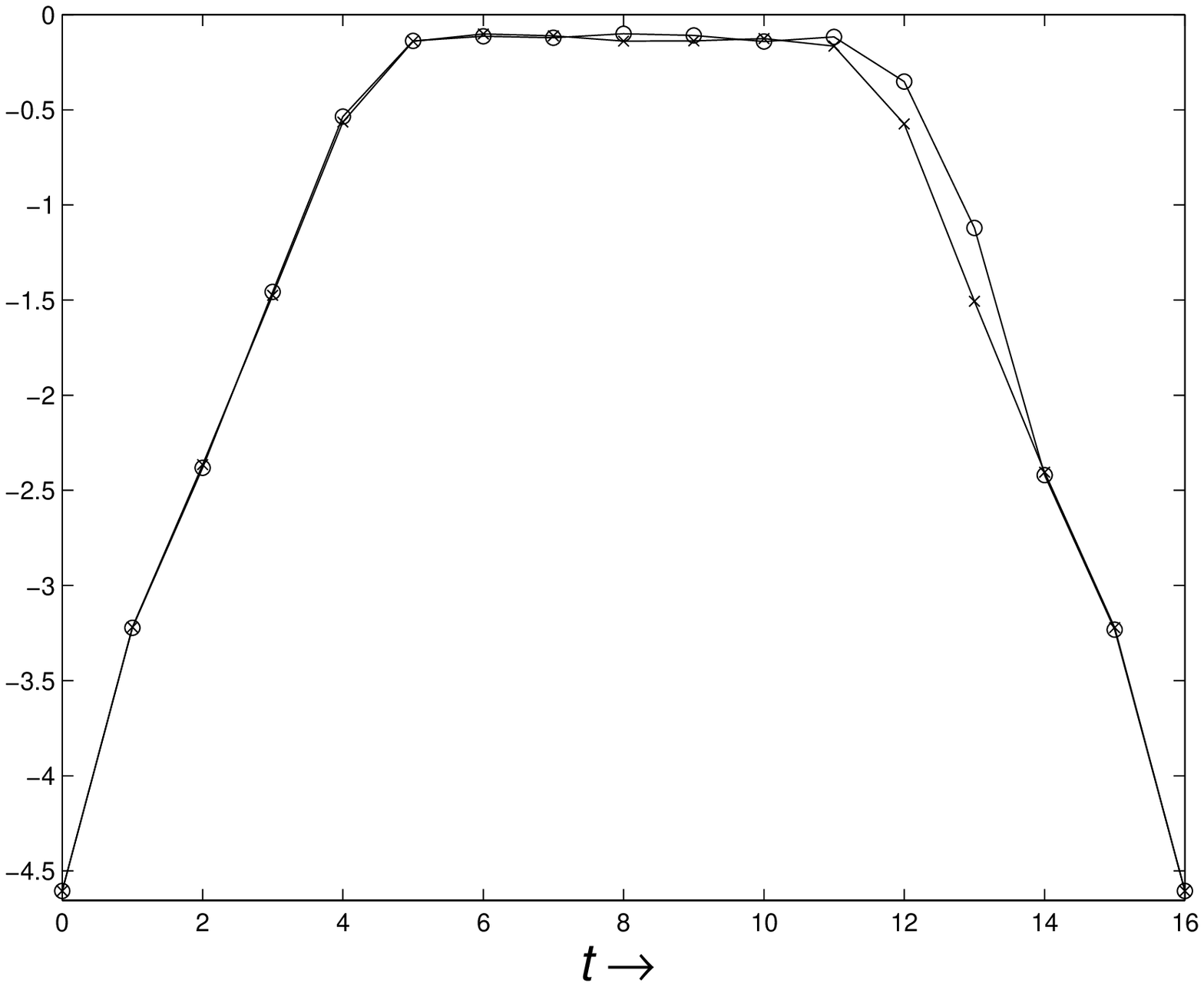}  }  
\endgroup
\NI Figure 1. Entropy, $S$, as a function of time for a mixing system, with two-time conditioning. For the left figure (2a) there is no perturbation. In the middle (2b) there is a perturbation at time 3.  On the right (2c) the perturbation is nominally at time 14, although because of the way entropy is calculated ({\it after} a time step, in terms of the nonthermodynamic parameter $t$) it is effectively at time 13$\half$.
\vskip 1 pt
\hbox to \hsize {\vrule  width 6 true in height .4pt depth 0pt}
\endinsert

\def\catmap{\phi_{\hbox{\romsix C}}}

In Fig.\ 1 I show the effect of using two-time boundary values on the dynamics of the cat map, $\catmap$. (This is a map of the unit square into itself with the rule: $x'\equiv x+y$, $y' \equiv x+2y$, mod~1. It is a mixing transformation, intensively exploited in ergodic theory [\arnold\rk] and I have used it as an example for two-time boundary value problems in many places, [\timebook\rk], etc.)

On the left (1a) there is no perturbation. The boundary conditions are that the system must be in a particular coarse grain (of size $0.1\!\times\!0.1$) at times 0 and 16. Evidently the entropy is a more or less symmetric function of time. (The statistical error comes from using a sample of 500 points, rather than the set $\epsilon$.) 

In terms of the direction of entropy increase it is that clear this arrow is a consequence of the boundary values given.

In the middle figure (1b), a perturbation is applied at time-3. At that time, instead of $\catmap$, a different map is applied ($x'\equiv 2x+3y$, $y' \equiv x+2y$, mod~1; this is {\it more} chaotic, and equilibrates faster than $\catmap$). The entropy is calculated (in the computer simulation) after time 3, ``after" in the sense of the nonthermodynamic parameter $t$, but reports this as the time-3 entropy. Because of this convention one can think of the perturbation as taking place at time-2$\half$. The deviation between the perturbed and unperturbed entropy is for times 3 and 4 (by time-5 both systems are in equilibrium). Because the perturbed system received a bigger kick at time-3 its entropy increases more rapidly.

The point of this figure is that the difference between the curves is confined to times later than the perturbation. The system shows causality.

The right hand figure (1c) shows a system that is perturbed at time-14. As before, the entropy-calculating convention makes this effectively a perturbation at time 13$\half$. In this case, the difference between perturbed and unperturbed systems is {\it before} the perturbation, ``before" in the sense of the nonthermodynamic parameter $t$. However, in terms of the direction of entropy increase, the entropy increase arrow and the causality arrow agree. This could be called reverse causality, but it is just normal causality with a bad choice of nonthermodynamic time parameter.

\topinsert
\begingroup 
\def\moveleftstart{-.2truein} 
\def\downskip{-.4truein} 
\def\moveleftmid{-.012truein} 
\def\figsize{1.8}
~\vskip \downskip
\hbox{\hskip\moveleftstart
\lsfig{\figsize}{\figsize}{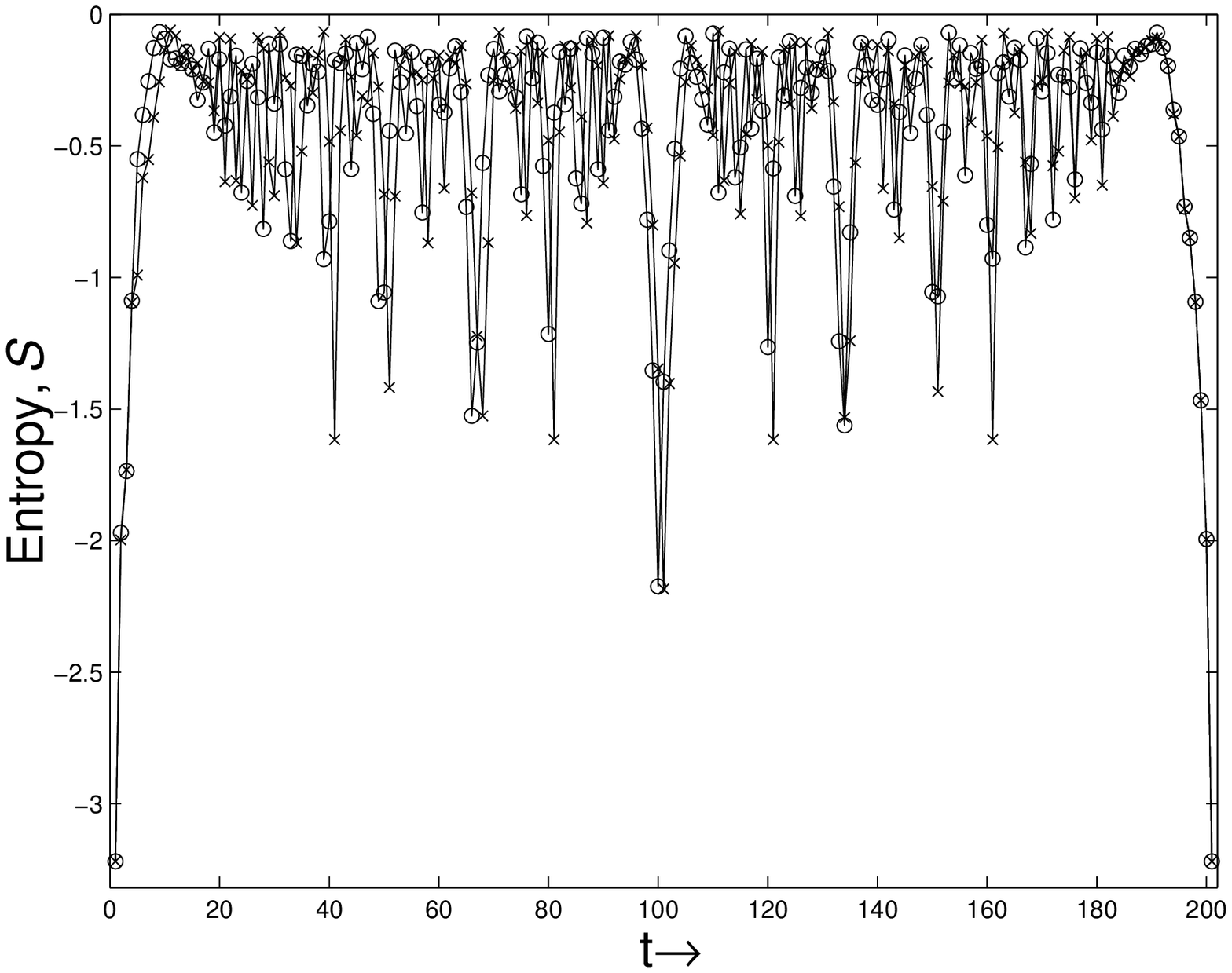}      \hskip \moveleftmid
\lsfig{\figsize}{\figsize}{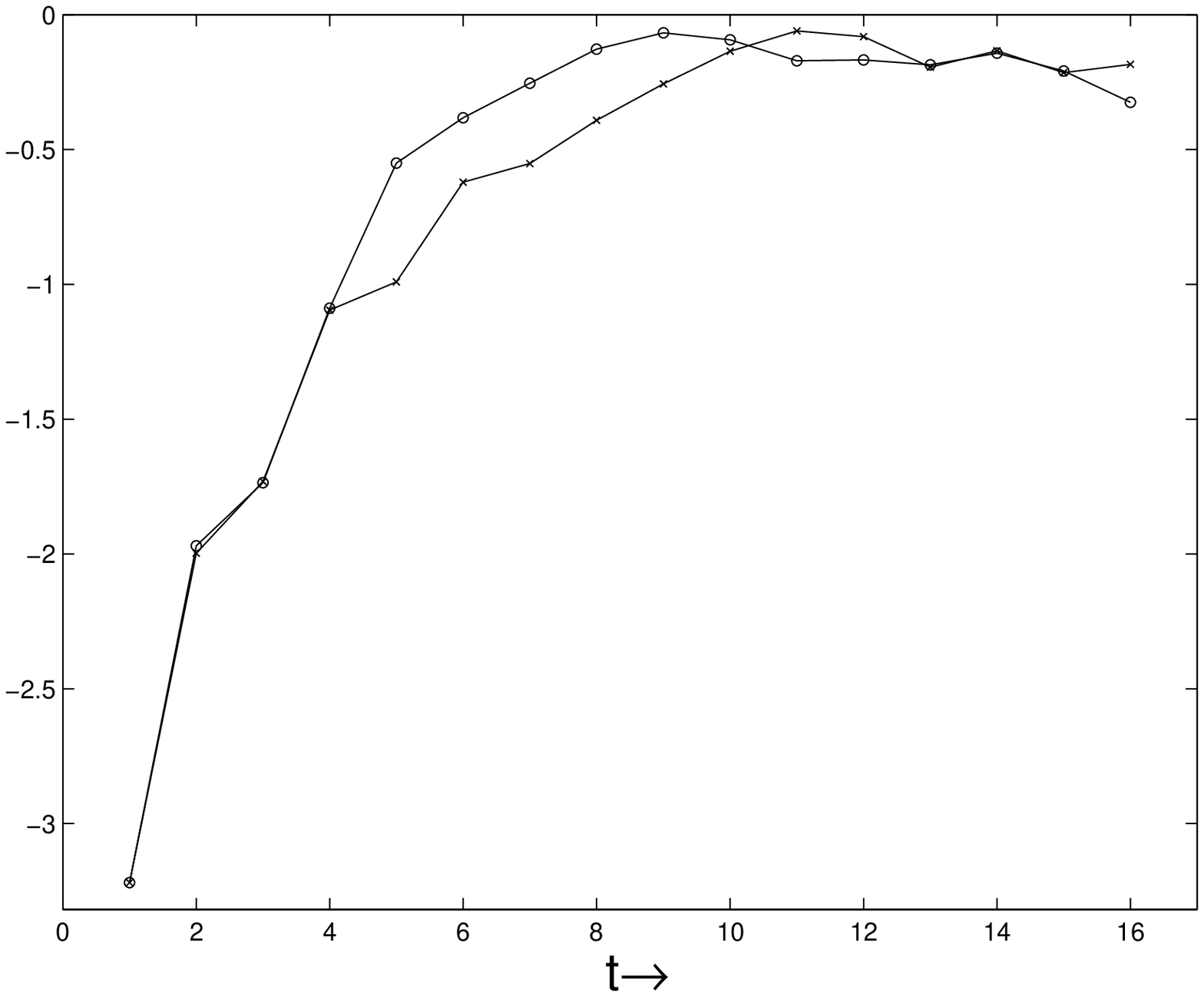}   \hskip \moveleftmid
\lsfig{\figsize}{\figsize}{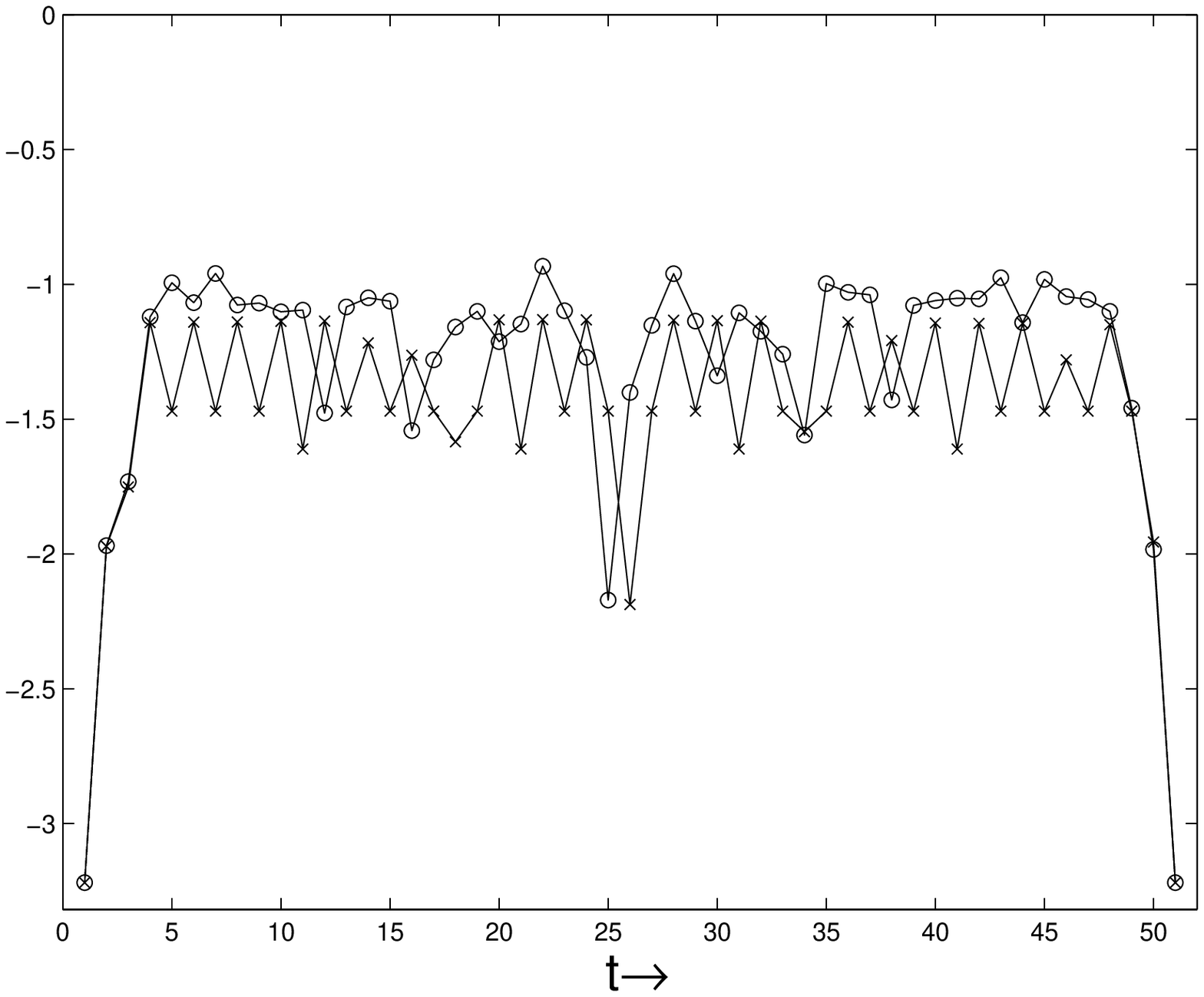}  }  
\endgroup
\NI Figure 2. Entropy, $S$, as a function of time, for oscillators with two-time conditioning. The left figure (2a) shows an entire run of 200 time steps. Both perturbed and unperturbed motion appear. They differ in many places. Note also the half-time depression in entropy, as well as other, smaller reductions. In the middle (2b) is shown only the first few time steps. The perturbation is at time-4. Causality is evident.  On the right (2c) only 50 time steps are used. Although from the previous figures it is clear that the entropy can reach its maximum values within about 10 time steps, when the conditioning time is 50 (as in 2c) the system cannot get near the equilibrium value. (All figures show the same numerical range of entropy.)
\vskip 1 pt
\hbox to \hsize {\vrule  width 6 true in height .4pt depth 0pt}
\endinsert

Finally I show what happens for harmonic oscillators. The perturbation is slightly different from that studied analytically above. Rather than a displacement, the system advances by $3\nu$ instead of $\nu$. The results are essentially the same and by this small change one also can see a level of robustness of the phenomenon.

For Fig.\ 2 there are 25 coarse grains along the ``$x$" direction and the frequency interval is of width $1/10$. Thus unconditioned relaxation should take place in about 10 time steps, but full equilibration should take about 250. For Fig.\ 2a and 2b (which are from the same run) both aspects are evident. With 200 time steps the system does approach $S=0$ and the relaxation time is about 10 (as is seen more easily in Fig.\ 2b). On the other hand, in Fig.\ 2c, with conditioning for 50 time steps, $S$ is far from 0, so that the potential for 10-time-step relaxation is thwarted by the future condition. In all cases there is a perturbation at time-4. Fig.\ 2b clearly shows that there is causality in this case. On the other hand, for the third figure, the system's relaxation is so compromised that the action of the perturbation takes place when the system has reached its maximum, although reduced, entropy level.

\header{Acknowledgements}
I am grateful to the Istituto Italiano per gli Studi Filosofici, Naples, for hosting the conference where this material was presented. My research is supported in part by the United States National Science Foundation grant PHY 97 21459.

\header{Appendix A. Entropy increase, stochastic dynamics and coarse graining}

The formalism developed above is useful for a general derivation of entropy nondecrease. The derivation also holds for quantum mechanics.\newline
{\it Proposition:}~ Coarse graining a distribution function, evolving it forward, and then again coarse graining, either increases the entropy or leaves it unchanged.

Let the distribution function for a classical system at time-0 be $\rho$. It is coarse grained to yield $\widehat\rho$, which is taken as $\rho(0)$. Thus
$$\rho(0)=\sum \frac{\chi_\alpha}{v_\alpha} \rho_\alpha\,
        \qquad\hbox{with}\quad \rho_\alpha=\int\chi_\alpha\rho \,.$$
From Sec.\ 2, the entropy of this distribution is $S(\rho_\alpha|v_\alpha)= -\sum \rho_\alpha \log(\rho_\alpha/v_\alpha)$. (Recall that $v_\alpha= \mu(\Delta_\alpha)$, the volume of coarse grain $\alpha$.) At time-$t$, $\widehat\rho$ becomes
$$\widehat\rho(t)=\sum \frac{\chi_{\tilde\alpha}}{v_\alpha}\rho_\alpha$$
with $\chi_{\tilde\alpha}$ the characteristic function of $\tilde\alpha \equiv \ph{t} (\Delta_\alpha)$. Now coarse grain again. This is the step where entropy nondecrease is forced, and I discuss its physical significance below. Coarse graining the function $\chi_{\tilde\alpha}$, the distribution function becomes
$$ \rho(t)=\widehat{\widehat\rho(t)}=
  \sum_{\alpha,\beta}\chi_\beta \rho_\alpha
         \frac1{v_\alpha v_\beta}\mu\left(\Delta_\beta\cap\ph{t}(\Delta_\alpha)\right) 
    =\sum \frac{\chi_\beta}{v_\beta} \rho'_\beta\,,  $$
with
$$  \rho'_\beta=\sum_\alpha R(\beta,\alpha)\rho_\alpha
         \quad\hbox{and}\quad
   R(\beta,\alpha)\equiv\mu\left(\Delta_\beta\cap\ph{t}(\Delta_\alpha)\right)\Big/v_\alpha \,.$$
It follows that the entropy of the distribution $\rho'$ is $S\left((R\rho)_\beta | v_\beta\right)$. Thus to establish the proposition above I must show that $S(R\rho|v)\geq S(\rho|v)$.

First I show that the matrix $R$ is {\it stochastic}, i.e., its elements are nonnegative and each column sums to one. The sum is $\sum_\beta R(\beta,\alpha) =  \mu\left[(\cup_\beta \Delta_\beta) \cup\ph{t}(\Delta_\alpha)\right]/v_\alpha$. Since $\ph{t}$ is measure preserving this sum gives unity. Furthermore, $Rv=v$, with $v$ the vector of grain volumes.

It is a theorem [\nr\cover,\nr\master\rk] that for any pair of distributions, $p$ and $q$, for which $S(p|q)$ is defined, and for any stochastic matrix $M$, $0\geq S(Mp|Mq)\geq S(p|q)$. Apply this to $\rho$ and $\rho'$. Making use of $Rv=v$, the proposition stated above on entropy nondecrease is established for classical dynamics.

The physical content of this derivation was incorporated in the replacement of $\widehat\rho(t)$ by its coarse grained smearing. The assumption is that {\it within each grain} the phase space points have spread uniformly. Thus for physical application of this proposition $t$ cannot be arbitrarily small. It must exceed a microscopic relaxation time associated with the coarse grains. Moreover, in coarse graining there is a destruction of information---monotonic entropy behavior contradicts the entropy reversal one gets using low-entropy two-time boundary conditions, as well as the more traditional counterexamples arising from Poincar\'e recurrence and time reversal.

\topinsert 
\begingroup \par \baselineskip 10pt \romnine \def \it{\italnine} \def\bf{boldnine}
\NI
 Table 1. Classical-quantum correspondence for the entropy increase proposition.
\medskip
\settabs 10 \columns
\+ $\Omega$        & $\cal H$ (Hilbert space)
      &&&& $\rho$     &  $\rho$ (density matrix)       \cr
\+ $\Delta_\alpha$ & ${\cal H}_\alpha$ (subspace) 
      &&&& $\mu$  &  Trace                            \cr
\+ $\chi_\alpha$   & $P_\alpha$ (projector)       
      &&&& $v_\alpha$  & dimension of ${\cal H}_\alpha$   \cr
\+$\ph{t}$   & $U_t$ \cr
\endgroup
\vskip -16pt \newline \NI 
\hbox to \hsize {\vrule  width \hsize height .4pt depth 0pt}
 \endinsert

The quantum version of this proposition involves no new mathematics, only a correspondence between the classical and quantum quantities. See Table 1. I find
$$\widehat\rho=\sum \frac{P_\alpha}{v_\alpha}\rho_\alpha\,, \hbox{~~with~}\rho_\alpha=\Tr P_\alpha \rho \hbox{~~and~} v_\alpha=\Tr P_\alpha  \,.$$
Entropy is again $S(\rho_\alpha|v_\alpha)$. (The $v$s no longer sum to unity, but this makes no essential difference.) Time evolution is given by a unitary operator, $U_t$, acting in the usual way: $\rho(t)=U_t \rho(0) U^\dagger_t$. Carrying through the same steps as for the classical case, coarse graining, evolving in time and coarse graining again, leads to the same equations, but with the matrix ``$R$" now given by
$$ R(\beta,\alpha)=\Tr \left[P_\beta U_t P_\alpha U^\dagger_t \right]\Big/v_\alpha \,.$$
Stochasiticity of $R$ is readily established and entropy nondecrease follows as above.

\header{Appendix B. Perturbation in a deterministic system}

A perturbation is often thought of as an act of control. In contrast, it would seem that imposing future conditions denies the possibility of modified evolution. Put differently, perturbing is an act of free will; future conditions---along with the deterministic context for their imposition---fly in the face of that concept.

This is not the place for a discussion of free will, except to mention that contrary to the impression of many physicists, some philosophers find justification for free will, not from the supposed indeterminism of quantum mechanics, but from chaos in deterministic dynamical systems [\nr\dennet\rk, p.\ 152].

But one need not imagine an independent actor to obtain the ``perturbation" of Sec.\ 4. Consider the following situation, within the context of the cosmological scenario described in [\timebook] or [\oppvail\rk]. Two systems, A and B, are small parts of a big universe, but they are isolated, or nearly so, between the times to be used for the boundary value problem. The actual macroscopic boundary values for the two of them are the same. Now imagine that one of them, say B, is not perfectly isolated, but at some intermediate time, $t_0$, in its history, is struck by something coming in from the outside. This ``outside" is simply another part of the universe, not A and not B. Its main properties are its lack of correlation with what is otherwise happening to A and B, and its ability to pack a macroscopic wallop in B. Despite the outside force, I still require the same boundary values for A and B.

Now compare the macroscopic motions of A and of B. Were it not for the outside force, they should be the same. With the force, having changes occur {\it only} on one (temporal) side of the perturbation is what I call macroscopic causality.

\header{References}

\pritem{\newton} R. G. Newton, {\it Thinking about Physics} (Princeton University Press, Princeton, 2000).

\pritem{\gold} T. Gold, The Arrow of Time, Am.\ J. Phys.\ {\bf 30}, 403 (1962).

\pritem{\correlating} L. S. Schulman, Correlating Arrows of Time, Phys.\ Rev.\ D {\bf 7}, 2868 (1973).

\pritem{\timebook} L. S. Schulman, {\it Time's Arrows and Quantum Measurement} (Cambridge University Press, Cambridge, 1997).

\pritem{\opposite} L. S. Schulman, Opposite Thermodynamic Arrows of Time, Phys.\ Rev.\ Lett.\ {\bf 83}, 5419 (1999) (cond-mat/9911101).

\pritem{\oppvail} L. S. Schulman, A compromised arrow of time, to appear in Proc.\ of {\it \'Equations aux D\'eriv\'ees Partielles et Physique Math\'ematique}, Paris, June 2000, ed.\ B. Gaveau et al.\ (cond-mat/0009139).

\pritem{\arnold} V. I. Arnold and A. Avez,  {\it Ergodic Problems of Classical Mechanics} (Benjamin, New York, 1968).

\pritem{\differentdef} This definition differs slightly from that used in [\timebook\rk], etc. The difference is $-\sum v_\alpha \log v_\alpha$. Thus the maximum (old definition) entropy for $G$ equal-volume coarse grains is $\log G$. With the definition here the maximum is zero.

\pritem{\accuracy} L. S. Schulman, Accuracy of the semiclassical approximation for the time dependent propagator, J. Phys.\ A {\bf 27}, 1703 (1994).

\pritem{\goodaccuracy} I will not try to look for minimal values of $\tau$, since keeping track of this would necessarily include dependence on which sets ($\epsilon$s) are considered, as well as on the coarse grains. For any given collection of grains and boundary conditions, the mixing property guarantees that a $\tau$ can be found. In simple systems and with the $\Delta$s and $\epsilon$s all about the same size, $\tau \sim -\log\mu(\Delta)$.

\pritem{\unstable} L. S. Schulman, Time Reversal for Unstable Particles, Ann.\ Phys.\ {\bf 72}, 489 (1972).

\pritem{\huerta} H. S. Robertson, {\it Statistical Thermophysics} (Prentice-Hall,
 Englewood Cliffs, New Jersey 1993); M. A. Huerta and H. S. Robertson, Entropy, Information Theory, and the Approach to Equilibrium of Coupled Harmonic Oscillator Systems, J. Stat.\ Phys.\ {\bf 1}, 393 (1969);  M. A. Huerta and H. S. Robertson, Approach to Equilibrium of Coupled Harmonic Oscillator Systems. II, J. Stat.\ Phys.\ {\bf 3}, 171 (1971).

\pritem{\mixingpaper} In earlier work an arrow was derived from an asymmetric, dissipative perturbation, rather than from proximity to one or another boundary-value-induced low entropy state. See L. S. Schulman and R. Shtokhamer, Thermodynamic Arrow for a Mixing System, Int.\ J. Theor.\ Phys.\ {\bf 16}, 287 (1977).

\pritem{\onlymeaningful} There is no point in taking $t_0$ in the time interval in which the system has equilibrated. Since the perturbation is nondissipative, it will have no impact in either direction of time (i.e., the system stays in equilibrium).


\pritem{\cover} T. M. Cover and J. A. Thomas, {\it Elements of Information Theory} (Wiley, New York, 1991).

\pritem{\master} B. Gaveau and L. S. Schulman, Master equation based formulation of non-equilibrium statistical mechanics, J. Math.\ Phys.\ {\bf 37}, 3897 (1996).

\pritem{\dennet} D. C. Dennett, {\it Elbow Room: The Varieties of Free Will Worth Wanting\/} (MIT Press, Cambridge, Mass., 1984).

\end